\documentclass[preprint,showpacs,aps]{revtex4}

\usepackage{graphicx}
\usepackage{dcolumn}
\usepackage{bm}

\begin{document}

\begin{titlepage}

\title{Suppression of spin-polarization in graphene nanoribbon by edge defect and impurity}

\author{Bing Huang$^1$, Feng Liu$^2$\footnote{Email: fliu@eng.utah.edu}, Jian Wu$^1$, Bing-Lin Gu$^1$, and Wenhui Duan$^1$\footnote{Email: dwh@phys.tsinghua.edu.cn}}

\address{$^1$Department of Physics, Tsinghua University, Beijing 100084, PRC}
\address{$^2$Department of Materials Science and Engineering, University of Utah, Salt Lake City, Utah 84112}

\date{\today}

\begin{abstract}
We investigate the effect of edge defects (vacancies) and
impurities (substitutional dopants) on the robustness of
spin-polarization in graphene nanoribbons (GNRs) with zigzag
edges, using density-functional-theory calculations. The stability
of the spin state and its magnetic moments is found to decrease
continuously with increasing concentration of defects or
impurities. The system generally becomes non-magnetic at the
concentration of one edge defect (impurity) per $\sim$ 10 \AA. The
spin suppression is shown to be caused by reduction and removal of
edge states at the Fermi energy. Our analysis implies an important
criterion on the GNR samples for spintronics applications.
\end{abstract}


\maketitle

 \draft

\vspace{2mm}

\end{titlepage}
Graphene nanoribbons (GNRs) have attracted much recent interest
because of their unique electronic properties and potential for
device applications \cite{Nakada, Wakabayashi, Kusakabe, Lee,
Louie, T. B. Martins, Qimin, Pisani, Kim, Hod, White}. Of
particular interest are those GNRs with zigzag edges, which are
shown to have a spin-polarized ground state \cite{Louie, Pisani,
Lee}. The spin-polarization is originated from the edge states
that introduce a high density of state at the Fermi energy. It can
be qualitatively understood in terms of Stoner magnetism of {\it
sp} electrons (in analogy to conventional {\it d} electrons)
occupying a very narrow edge band to render an instability of
spin-band splitting \cite{D. M. Edwards}. The spins are found to
be localized on the ribbon edges, with ferromagnetic coupling
within each edge and antiferromagnetic coupling between the two
edges.

Because the magnetism in GNRs is resulted from the highly
degenerate edge states, it must require a perfect edge structure.
However, real samples of GNRs \cite{Kim} are unlikely to have
perfect edges but contain structural defects and impurities of
foreign atomic species. Thus, one important question is how robust
is the spin state in presence of edge defects and impurities. The
answer to this question not only is scientifically interesting to
better understand the physical mechanisms of spin-polarization in
GNRs, but also have important technologically implications if GNRs
are to be realized as a new class of spintronic materials. Some
existing studies \cite{Louie} seemed to suggest that the spins in
GNRs are robust against formation of edge defects at certain
concentrations. However, in this Letter, we show that the spins
can be completely removed by edge defects or impurities at
concentrations accessible in real samples, which renders a
practical difficult for realizing spin-polarization in GNRs.

We have performed a systematic study of the stability and degree
of spin-polarization in GNRs as a function of the concentration of
edge defects and impurities, using first-principles
density-functional-theory (DFT) calculations, We chose vacancy and
substitutional boron (B) atom, as typical examples of structural
edge defect and impurity, respectively, since vacancies are
expected to be the most abundant defects in a roughened edge and B
atoms are a common choice of electronic dopants \cite{Qimin, T. B.
Martins}. Our calculations show that the spin is completely
suppressed at the defect (impurity) sites, while it may preserve
on those sites away from defects (impurities) when the defect
(impurity) concentration is low. The stability of the spin state
and its "average" magnetic moment decreases continuously with
increasing defect (impurity) concentration. Typically, independent
of ribbon width, the system becomes non-magnetic at the critical
concentration of one edge defect (impurity) per $\sim$ 10 \AA, a
concentration shown to be accessible in real samples based on
thermodynamic considerations. The magnetism can also be "dead"
locally in between two defects (impurities) as long as they are
closer than the third nearest-neighbor (NN) positions. The spin
suppression correlates closely with the reduction and removal of
edge states at the Fermi energy.

\begin{figure}[tbp]
\includegraphics[width=8.5cm]{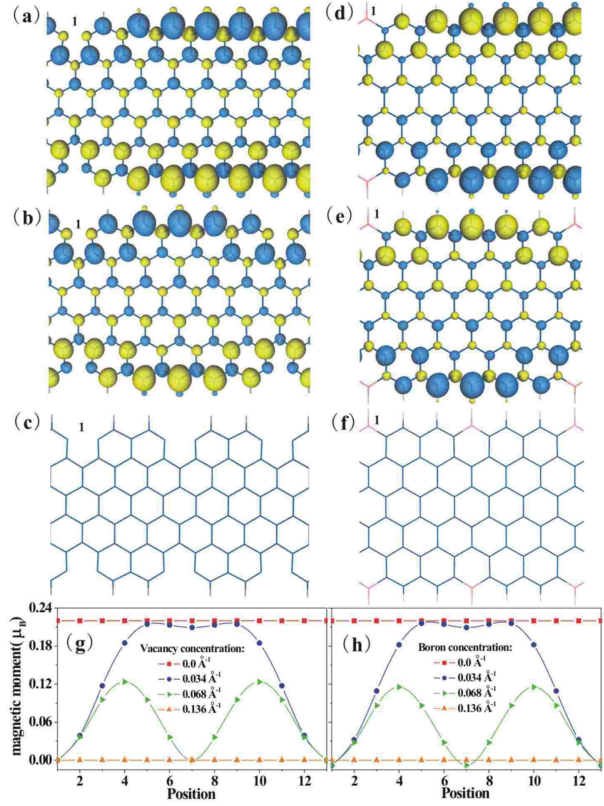}
\caption{ Isovalue surfaces of charge difference between the
spin-up and spin-down states of the ground states calculated for
the (2.5,2.5) GNR at three different linear vacancy concentrations
on the edge: (a)0.034 \AA$^{-1}$, (b)0.068 \AA$^{-1}$, and
(c)0.136 \AA$^{-1}$. Half of the supercell in Fig. 1a, one in Fi.g
1b and two in Fig. 1c are shown. The yellow and blue surface
represents spin-up and spin-down state, respectively. The range of
isovalues is set at [-0.005:0.005] $\mu_{B}$ \AA$^{-1}$ in case
(a) and [-0.003:0.003] $\mu_{B}$ \AA$^{-1}$ in case of (b) and
(c). (d)-(f) The same as (a)-(c) for B impurities. (g) The
magnetic moment per edge atom on the top edge, at three vacancy
concentrations corresponding to the case of (a), (b) and (c), plus
the case without vacancy. (h)The same as (g) for B impurities.}
\end{figure}

Our calculations are performed using DFT pseudopotential
plane-wave method within the local spin density
approximation\cite{VASP}. The plane wave cut-off energy is set as
450 eV. The structure optimization has been performed until the
residual atomic forces are less than 0.01 eV/\AA. We calculated
GNRs with H-terminated zigzag edges of two different widths,
namely the (2.5,2.5) and (1.5,1.5) "armchair" ribbons
\cite{Qimin}, using a nomenclature in analogy to armchair carbon
nanotubes that would unfold into corresponding ribbons with zigzag
edges. Vacancies or impurities are introduced in the edge by
removing edge C-H pairs or replacing C with B, respectively. Their
concentrations are varied from 0.034 \AA$^{-1}$ to 0.136
\AA$^{-1}$ along the edge, and both uniform and non-uniform
distributions of defects (impurities) are considered.

Figure 1a shows the calculated ground-state spin distributions of
a (2.5,2.5) GNR at an edge vacancy concentration of ~0.034
\AA$^{-1}$. The spin polarization is mostly localized on the edges
with the top and bottom edge spins anti-ferromagnetic coupled with
each other. One notices that spin is locally absent at the vacancy
sites and it increases as one moves away from the vacancy along
the edge. Figure 1b shows the ground-state spin distributions at a
higher vacancy concentration of ~0.068 \AA$^{-1}$. Overall, the
spins exhibit an identical distribution pattern, but their
magnitude is reduced in comparison to Fig. 1a. At even higher
vacancy concentration of ~0.136 \AA$^{-1}$, Figure 1c shows that
the spins are completely suppressed at all atomic sites and the
GNR becomes non-magnetic. Almost identical results are obtained
for B impurities as shown in Figs. 1d-1f.

In Fig. 1g, we plot the magnetic moments on the top edge as a
function of edge positions, at four different vacancy
concentrations. Without vacancy, there is a uniform spin
distribution with a large moment of $~0.22 \mu_B$ at every edge
sites (red squares). When vacancies are introduced at a low enough
concentration so that spins preserve on the edges, the presence of
vacancies in effect enforces a spin-density wave (oscillation)
along the ribbon edge, with the zero moment (nodal point) at the
vacancy site and the maximum moment in the middle between the
vacancies (blue dots and green right triangles). The period of the
spin density wave equals the inverse of vacancy concentration,
while its amplitude decreases with increasing concentration. At a
sufficiently high concentration, the vacancies suppress all the
spins and the moments become zero everywhere (orange up triangle).
The same results are obtained for B impurities, as shown in Fig.
1h.

\begin{figure}[tbp]
\includegraphics[width=8.0cm]{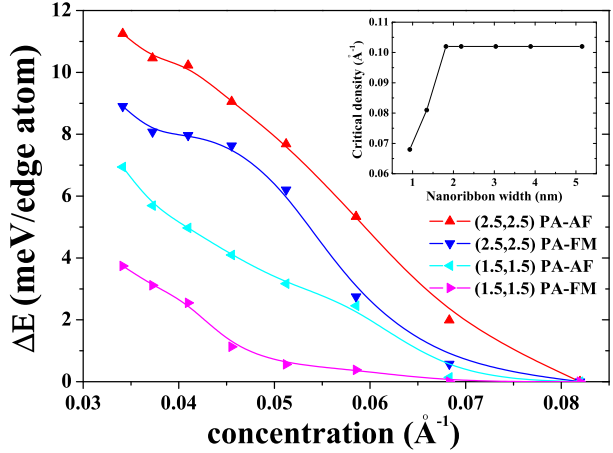}
\caption{The energy difference per edge atom between the magnetic
(AF or FM) and paramagnetic (PA) state as a function of vacancy
concentration in the edge. Two different ribbon widths of (2.5,
2.5) and (1.5,1.5) are shown. The inset shows the critical
concentration as a function of ribbon width up to 5 nm.}
\end{figure}

The results in Fig. 1 indicate that the spin-polarization
decreases with increasing edge defect (impurity) concentration and
eventually vanishes. To further demonstrate this phenomenon, we
have compared the relative stability of the spin state with the
non-spin state. Figure 2 shows the calculated energy difference
per edge atom as a function of vacancy concentration. The results
of B impurities are essentially the same. In general, the energy
difference between the magnetic state (both AF and FM) and the
nonmagnetic state decreases rapidly with increasing defect
(impurity) concentration. Eventually, the difference decreases to
zero and the GNR becomes nonmagnetic. We have examined the
critical defect (impurity) concentration at which the spin
vanishes as a function of ribbon width from 1 to 5 nm, as shown in
the inset of Fig. 2 (Note the curve is discontinuous at the point
of 2 nm because we can only change ribbon width and concentration
discretely.). We found that the spin-polarization is completely
suppressed at a typical critical defect (impurity) concentration
of $\sim$ 0.10 \AA$^{-1}$ when the ribbon width is larger than 2
nm. The critical concentration is lower in narrower ribbons, e.g.,
at ~0.081 \AA$^{-1}$ for a (2.5,2.5) ribbon and at ~0.068
\AA$^{-1}$ for a (1.5,1.5) ribbon.

Given the critical concentration of $0.10 \AA^{-1}$, the next
important question is whether such a concentration is accessible
in real samples. Thus, we have calculated the formation energies
of edge defects and impurities. For vacancy formation, we obtain
$\Delta E = 9.21 eV + \mu_C$ for removing a C away from the edge,
where $\mu_C$ is the chemical potential of C which may vary from
-10.13 eV (bulk graphene) to -1.25 eV (atomic C). Now, if we
consider the vacancy to be thermally activated, then at room
temperature the critical concentration of $0.10 \AA^{-1}$ can be
achieved by a formation energy of 0.035 eV, which requires a C
chemical potential of -9.175 eV within the possible range of C
chemical potential variation. In addition to thermal excitation,
vacancies or structural defects are also forced by nanopatterning
process. In fact, presently the ribbons can only be made with very
rough edges containing a high concentration of defects \cite{Kim}.
Thus, the critical density we identify here is likely to be
accessible in real samples. This will likely pose a practical
challenge in realizing the spin-polarization in GNRs. For the case
of B impurity, we obtain $\Delta E = 2.49 eV + \mu_C - \mu_B$,
where the chemical potential of B may vary from -7.49 eV
($\alpha$-phase B) to -0.31 eV (atomic B). In this case, the
critical concentration of B is accessible by increasing B partial
pressure and hence its chemical potential in the doping process.

\begin{figure}[tbp]
\includegraphics[width=8.0cm]{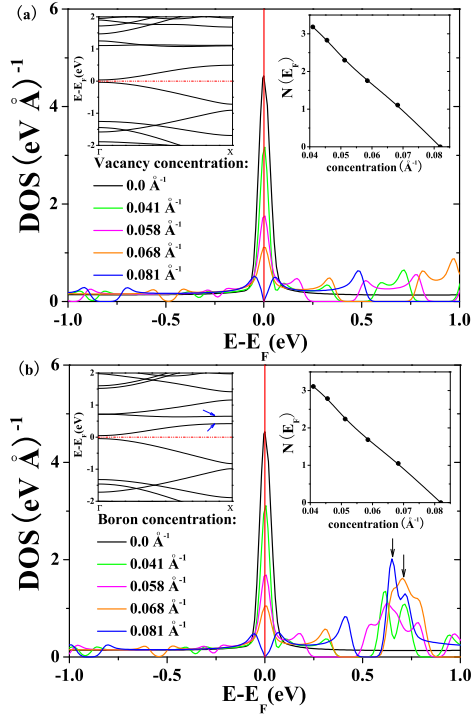}
\caption{(a)The calculated total DOS for a (2.5,2.5) ribbon in the
paramagnetic state at five different vacancy edge concentration.
Up-left inset: the band structure at the vacancy concentration of
0.081 \AA$^{-1}$. Up-right inset: the DOS at $E_{F}$ as a function
of vacancy concentration. (b) The same as (a) for B impurities.}
\end{figure}

The magnetism is resulted from a high density of state (DOS) at
the Fermi Energy ($E_F$) that renders an instability of spin
polarization. Consequently, one expects the suppression of
magnetism is correlated with the change of DOS at $E_F$ induced by
defects (impurities). We have examined the DOS as a function of
defect (impurity) concentration, which has indeed confirmed such
mechanism. In Fig. 3a, we plot the DOS in the paramagnetic state
of the (2.5,2.5) ribbon as a function of vacancy concentration.
Clearly, the DOS at $E_{F}$ decreases almost linearly (up-right
inset) with increasing vacancy concentration, in close correlation
with the magnetization shown in Figs. 1 and 2. As the defect
concentration reaches 0.081 \AA$^{-1}$, the DOS at $E_{F}$ falls
to zero, and the system becomes non-magnetic. Without defects, the
highly degenerate edge states in a perfect ribbon edge give rise
to the high DOS at $E_{F}$. Since the vacancies will not
contribute to the same edge state, their presence decreases the
DOS at $E_{F}$.

Similarly, in Fig. 3b, we plot the DOS in the paramagnetic state
of the (2.5,2.5) ribbon as a function of B impurity concentration.
Again, the DOS at $E_{F}$ decreases almost linearly (up-right
inset) with increasing B concentration and eventually vanishes, so
that the ribbon becomes nonmagnetic. However, here the mechanism
of DOS reduction at $E_{F}$ is slightly different from the case of
vacancy. The impurity B atoms act as electronic dopants to
introduce impurity states (levels) away from the Fermi energy so
as to reduce the DOS at $E_{F}$. This can be seen in Fig. 3b when
the DOS at $E_{F}$ decreases and vanishes at $E=0$ with increasing
B concentration, another peak of DOS appears and increases at $E =
0.7 eV$. Those bands of edge states (up-left inset) being shifted
to 0.7 eV are indicated by arrows. So effectively, the B
impurities lift the degeneracy of edge states and hence decrease
the DOS at $E_{F}$. In addition, the corresponding band structure
(up-left inset) shows the system becomes semiconducting with a
small band gap upon B doping. (We have also studied the case of N
atoms, which showed the same effect as B atoms, except the
impurity edge states were shifted to below Fermi energy since N is
a {\it n}-type dopant.)

We may further understand the spin suppression by edge defect
(impurity) in the context of itinerant ferromagnetism and local
order \cite {Liu1,Liu2,Liu3}. It has been shown that the magnetic
moment in an itinerant magnetic materials depends strongly on
local coordination\cite {Liu1,Liu2}. When a nonmagnetic "impurity"
is introduced in the magnetic medium, the moment is greatly
suppressed at the impurity site and its vicinity\cite{Liu2}.
Conversely, when a magnetic element is introduced in a nonmagnetic
medium, its moment is quenched at low concentration but can be
redeveloped at high concentration and there is strong correlation
between the magnetic dopants\cite{Liu3}.

Similarly, our case here can be thought of doping the magnetic
medium of itinerant {\it sp} electrons \cite{D. M. Edwards} with
nonmagnetic impurities, vacancy or B atom, which will suppress the
magnetic moment at the impurity site and its vicinity. When two
and more impurities are introduced, the correlation between them
would determine the local state of moment (spin-polarization). To
test this idea, we have examined the local magnetic moment in
between two defects (or impurities) as a function of their
separation. From Fig. 4a to 4c, we see that the local moment in
between the two vacancies decreases as the two vacancies moving
closer and completely vanishes when they are at the third NN
position (Fig. 4c). The same result is also obtained at even lower
defect concentrations. This means spin suppression is a rather
localized effect. For example, we can selectively introduce
defects (impurities) in only one ribbon edge while keep the other
edge perfect. Then we will create a ferromagnetic order along one
ribbon edge while the other edge is nonmagnetic, as shown in Fig.
5, which could be useful for spintronics applications if only
ferromagnetism is desirable.

\begin{figure}[tbp]
\includegraphics[width=8.0cm]{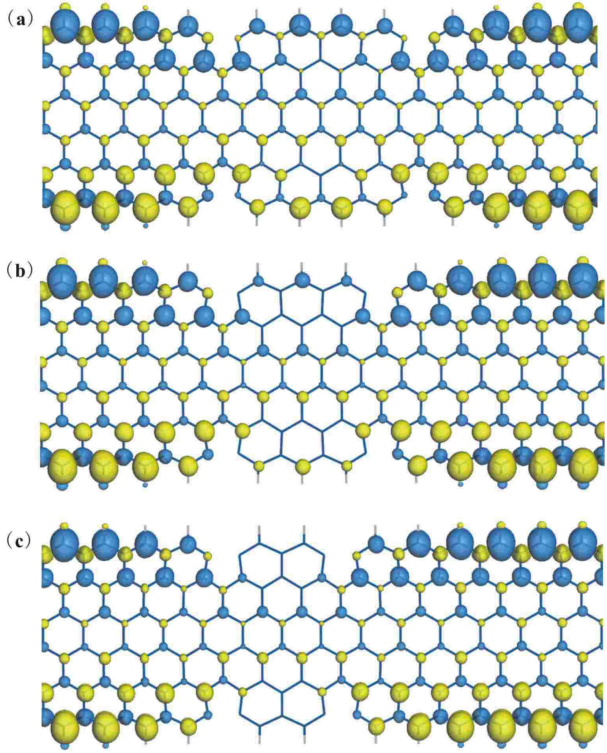}
\caption{Similar to Fig. 1a, but with the vacancies moving closer
from (a) the fifth NN position (even distribution) to (b) the
fourth NN position and to (c) the third NN position. The range of
isovalues is set at [-0.004:0.004] $\mu_{B}$ \AA$^{-1}$.}
\end{figure}

\begin{figure}[tbp]
\includegraphics[width=8.0cm]{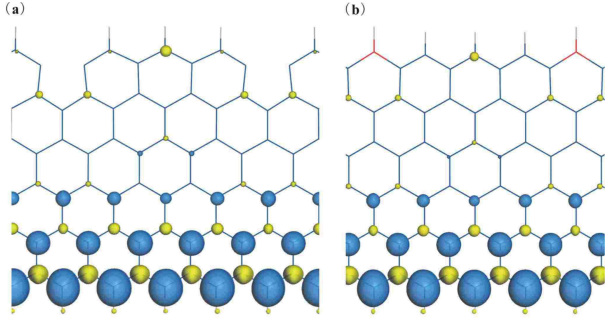}
\caption{Similar to Fig. 1a, but with (a) vacancy and (b) B
impurity introduced at only one of the two ribbon edges at the
fourth NN positions. The range of isovalues is set at
[-0.005:0.005] $\mu_{B}$ \AA$^{-1}$.}
\end{figure}

In conclusion, we have examined the robustness of
spin-polarization in GNRs with zigzag edge, using first principle
calculations. We show that the spin polarization can be greatly
suppressed in the presence of edge defects and impurities. In
general, the GNR becomes nonmagnetic at a critical edge defect
(impurity) concentration of $\sim$ 0.10 \AA$^{-1}$, which is shown
to be accessible in real samples. The magnetic moments may also
vanish locally if two defects (impurities) occur randomly to be
closer than the third NN distance. The spin suppression correlates
closely with the reduction of DOS at the Fermi energy induced by
defects (impurities). It can be qualitatively understood in the
context of itinerant magnetism and local order. Our findings
indicate that although the magnetic properties of GNRs are of
great scientific interest, practical realization of
spin-polarization in GNRs for spintronics applications can be
rather challenging. On one hand, the overall spin sate can be
completely removed by a high defect (impurity) concentration, and
on the other hand, the local spins can be suppressed by small
randomness of edge structure with a few concentrated defects,
which can be detrimental to "one-dimensional" spin transport along
the edge.

The work at Tsinghua is supported by the Ministry of Science and
Technology of China and NSFC, the work at Utah is supported by
DOE.

\newpage

\end{document}